\documentclass{article}
\usepackage{fullpage}

\newcommand{\ord}{{\cal O}}
\newcommand{\PP}{{\bf P}}
\newcommand{\NP}{{\bf NP}}
\newcommand{\NC}{{\bf NC}}
\newcommand{\CC}{{\bf CC}}

\begin{document}

\title{Computational complexity in physics}
\author{Cristopher Moore \\
Computer Science Department and Department of Physics and Astronomy \\
University of New Mexico, Albuquerque NM 87131 \\
and the Santa Fe Institute, Santa Fe, NM \\
{\tt moore@cs.unm.edu}}
\maketitle

There are many definitions of the word ``complexity.''  These fall
into three main categories:
\begin{itemize}
\item 
Information and thermodynamic entropy, as in Shannon and Gibbs, with
extensions by Crutchfield and others
\item
Minimum description length, as in Kolmogorov and Chaitin, and physical
versions of this, as in Lloyd, Pagels and Gell-Mann
\item 
Computational complexity.
\end{itemize}

More generally, computational complexity is about drawing qualitative
distinctions between problems based on how hard it is to solve them.
In a sense, computational complexity is descended from the concerns of
the constructivists on hearing from Cantor that some infinities are
bigger than others, e.g.\ that there are more subsets of the integers
than their are integers.  The constructivists denied that all subsets
of the integers have mathematical existence, and insisted that we
should only believe in those that have an effective description or
definition \cite{odifreddi}.  As the 20th century progressed, this
idea of a description came to mean an algorithm which could be carried
out by some mechanical device, such as a Turing machine, to tell
whether an integer is in the set or not.  The Church-Turing thesis
states that the Turing machine captures what is ``computable'' in any
reasonable way, and this has been borne out by the fact that it is
provably equivalent to lambda calculus, Boolean circuits, all known
programming languages, and every physically reasonable computing
device proposed to date.

As people actually started to build computers --- as Turing's ideas
were joined with von Neumann's --- the question became not just
whether a set or problem is computable or not, but whether it is {\em
feasibly} computable, i.e.\ with a reasonable amount of memory and in
a reasonable time.  As computers got faster and people solved larger
problems, the behavior of the computation time $f(n)$ in the limit of
large problem size $n$ became more and more important, causing us to
adopt more and more conservative definitions of what we mean by
feasible computation \cite{papa}.  In the 1950's, we asked how many
nested exponentials it took to define $f$, in the 70's and 80's we
asked whether $f$ is polynomial or not, and as massively parallel
computation becomes a reality we might demand that $f(n)$ be
polylogarithmic, i.e.\ $\ord(\log^k n)$ for some $k$, on a parallel
computer with a polynomial number of processors.

This last class, the set of efficiently parallelizable problems, is
called \NC; it is believed to be a proper subset of \PP, the
set of problems solvable in polynomial time.  A problem is {\em{\bf
P}-complete} if it is among the hardest problems in \PP, in the
sense that any other problem in {\PP} can be reduced to them.  {\bf
P}-complete problems are believed to be {\em inherently sequential},
meaning that a polynomial amount of work has to be done step-by-step
in order, and that even a massively parallel computer can't improve on
this very much --- just as having 10 cooks doesn't let you make dinner
in $1/10$ the time \cite{greenlaw}.

Note the analogy here to the distinction between chaotic systems,
which have to be integrated numerically, and integrable ones, which
can be predicted simply by plugging the initial conditions into a
closed-form solution.  For a chaotic system, the complexity of
predicting the system for a time $t$ in the future grows linearly with
$t$ (or even quadratically, given the increasing accuracy we need at
longer times) while for an integrable system this increases
polylogarithmically with $t$ (since $t$ has $\ord(\log t)$ digits that
need to be manipulated, and we assume that we can evaluate the
closed-form solution in time polynomial in the number of digits) or
not at all.

In fact, we can classify a number of interesting physical systems
based on whether the problem of predicting them is in {\PP} or {\NC}.
We can always predict a cellular automaton in polynomial time by
simulating it explicitly, just as we can numerically integrate a
differential equation.  Moreover, since cellular automata can easily
simulate Turing machines~\cite{mats}, this problem is \PP-complete in
general.  However, if the cellular automaton has certain algebraic
properties, we can predict it much more quickly, in $\ord(\log t)$ or
$\ord(\log^2 t)$ time.  Thus in special cases the CA prediction
problem can be in \NC, even for some nonlinear rules
\cite{quasi,semi}.

On the other hand, some systems such as lattice gases~\cite{lgas},
sandpiles~\cite{sand}, and zero-temperature Ising
dynamics~\cite{voting} are \PP-complete to predict, meaning that
unless $\PP=\NC$ (in which case all polynomial-time problems are
efficiently parallelizable) there is no way around simulating them
step-by-step.  This is because these systems' dynamics allow us to
build ``gadgets'' that carry information from place to place and
manipulate it logically, so that predicting the system is equivalent
to evaluating a Boolean circuit.  Interestingly, our proof for
sandpiles works in $d \ge 3$, and the complexity of sandpiles in two
dimensions (which is the case relevant to self-organized criticality
and conformal fields) remains an open question.

It is interesting to contemplate how this might be proved for systems
like the Navier-Stokes equations; it seems impossible that there are
short cuts for predicting hydrodynamics, but proving something like
\PP-completeness for a system like this seems very distant.  The
reason is that the only known technique for proving computational
hardness results is (essentially) to build a computer in the system,
like the Boolean circuits mentioned above.  Without a way to build a
computer entirely out of water --- whose only elements are the eddies
and vortices themselves --- we have no way to {\em prove} that the
system is hard.

In fact, if {\PP} and {\NC} are different it is known that there are
problems in between, outside {\NC} but not \PP-complete either.  It
may be that many physical systems lie in this gap, with neither a
clever fast algorithm for predicting them without simulation, nor a
way to build universal computers out of them.  Unfortunately, if this
is the case we have no idea how to prove that they are there.

An early application of computational complexity in physics was
Barahona's proof that finding the ground state of a spin glass is {\bf
NP}-complete \cite{barahona}, meaning that it is as hard as
optimization problems such as the Traveling Salesman's problem
\cite{garey}.  As far as we know, such problems take exponential time
to search through the space of possible solutions.  

Interestingly, two-dimensional spin glasses are in \PP, but adding
even one layer makes them {\NP}-complete.  Since this presumably does
not change their bulk properties, it is not clear how closely
computational complexity is related to physical properties.  The main
difference is that computational complexity is a worst-case notion,
while physics is concerned with the average case and bulk properties
of a material in the thermodynamic limit.  Some progress is being made
in extending computational complexity to the average
case~\cite{average}, but this is difficult since one must choose a
probability distribution for the notion of ``average'' to be
meaningful, and it is not always clear what the most natural
distribution is for a given problem.

In analogy to the random graphs of Erd\H{o}s and R\'enyi, where we
have a uniform distribution over all graphs with a given number of
nodes and a given number of edges, we can construct random problem
instances with a given number of variables and a given number of
constraints.  With this probability distribution, many {\NP}-complete
problems show a phase transition from satisfiability to
unsatisfiability as we increase the number of constraints per variable
beyond a critical threshold.  This has become an active area of
collaboration between computer scientists and physicists, including
rigorous upper and lower bounds on these transitions using
probabilistic arguments (e.g.\ \cite{dubois}, analysis of
polynomial-time heuristics (e.g.\ \cite{achsorkin}), and applications
of the replica trick (e.g.\ \cite{monasson}).

However, even for these phase transitions it is not clear to what
extent computational complexity and thermodynamic properties are
related.  An early conjecture was that {\NP}-complete problems have
first-order phase transitions, while problems in {\PP} have
second-order transitions.  However, some {\NP}-complete problems are
easy on average, and in the random case behave like problems in \PP.
For instance, 1-in-3 SAT is a variant of Satisfiability where we have
triplets of literals, each of which is a variable or a negated
variable, where we require exactly one literal in each triplet to be
true.  While 1-in-3 SAT is {\NP}-complete, its phase transition is in
the same universality class as that of 2-SAT, which is in \PP.  This
is because the {\NP}-completeness is buried under a simple percolation
process, and with probability 1 the problem is satisfiable if and only
if it is below the percolation threshold~\cite{soda}.

However, there does seem to be a connection between worst-case
complexity and physical properties fairly often.  For instance, Machta
showed that predicting diffusion-limited aggregation is {\bf
P}-complete, since its dynamics can simulate a NOR
gate~\cite{dla1,dla3}.  However, in the case of internal
diffusion-limited aggregation, also known as diffusion-limited
erosion, the dynamics are equivalent to a comparator gate, which
unlike the NOR is incapable of generating arbitrary Boolean functions.
This puts the worst-case complexity of IDLA in the (probably) lower
complexity class $\CC \subset \PP$.  In addition, a simple parallel
algorithm seems to work in polylogarithmic time, much faster than
explicit simulation, suggesting that predicting IDLA is in {\NC} on
average~\cite{idla}.  Physically, this seems to correspond to the fact
that DLA grows complex, dendritic clusters, while IDLA clusters are
spherical with logarithmic fluctuations.  This example suggests that
computational complexity really does say something about the dynamics
of a system, and how those dynamics allow it to propagate information
and enforce constraints between its degrees of freedom.

{\bf Acknowledgments.}  I am extremely grateful to Joseph McCauley,
Arne Skjeltorp, and the organizers of the NATO School on Complexity
and Large Deviations for inviting me to speak in Geilo.

\end{document}